\documentclass[aps,twocolumn,prl,floatfix,showpacs,preprintnumbers,amsmath,amssymb,nofootinbib,superscriptaddress]{revtex4-2}
\usepackage[normalem]{ulem}
\usepackage{wasysym}
\usepackage{color}
\usepackage{graphicx}
\usepackage[caption=false]{subfig}
\usepackage{dcolumn}    
\usepackage{multirow,booktabs}
\usepackage{comment}
\usepackage{soul}

\usepackage{physics, mathtools, braket}
\usepackage [ english ]{ babel }

\usepackage{amsmath,amsfonts,amssymb,amsthm} 
\usepackage{latexsym}
\usepackage{bbold}
\usepackage{feynmp-auto}

\usepackage{bigstrut}
\usepackage{amsmath,amsfonts,amsthm,bm}
\usepackage{CJK}
\usepackage[pdfstartview=FitH,
            CJKbookmarks=true,
            bookmarksnumbered=true,
            bookmarksopen=true,
            colorlinks,
            pdfborder=001,
            linkcolor=blue,
            anchorcolor=blue,
            citecolor=blue,
            urlcolor=blue,
            ]{hyperref}

\usepackage{orcidlink}

\graphicspath{{./}}
\newcommand{\Bl}{\mathcal{B}}
\newcommand{\Eop}{\mathcal{E}}
\newcommand{\Ox}{$^{16}$O}
\newcommand{\Ne}{$^{20}$Ne}
\newcommand{\vecalpha}{\vec{\alpha}}

\begin{document}


\title{Linking chiral nuclear forces to light-ion collisions: a global sensitivity analysis}

\author{Thomas Duguet}
\email{thomas.duguet@cea.fr}
\affiliation{IRFU, CEA, Universit\'e Paris-Saclay, 91191 Gif-sur-Yvette, France}

\author{Andreas Ekstr\"om \orcidlink{0000-0002-4917-4293}}
\email{andreas.ekstrom@chalmers.se}
\affiliation{Department of Physics and Astronomy, Chalmers University of Technology, SE-412 96 G\"oteborg, Sweden}

\author{Giuliano Giacalone \orcidlink{0000-0002-8776-1034}}
\email{giuliano.giacalone@cern.ch}
\affiliation{Theoretical Physics Department, CERN, CH-1211 Gen\`eve 23, Switzerland}

\author{\\Gaute Hagen \orcidlink{0000-0001-6019-1687}}
\email{hageng@ornl.gov}
\affiliation{Physics Division, Oak Ridge National Laboratory, Oak Ridge, Tennessee 37831, USA}
\affiliation{Department of Physics and Astronomy, University of Tennessee, Knoxville, Tennessee 37996, USA}
\affiliation{Department of Physics and Astronomy, Chalmers University of Technology, SE-412 96 G\"oteborg, Sweden}

\author{Takayuki Miyagi \orcidlink{0000-0002-6529-4164}}
\email{miyagi@nucl.ph.tsukuba.ac.jp}
\affiliation{Center for Computational Sciences, University of Tsukuba, 1-1-1 Tennodai, Tsukuba 305--8577, Japan}

\author{Alexander Tichai \orcidlink{0000-0002-0618-0685}}
\email{alexander.tichai@tu-darmstadt.de}
\affiliation{Technische Universit\"at Darmstadt, Department of Physics, 64289 Darmstadt, Germany}
\affiliation{ExtreMe Matter Institute EMMI, GSI Helmholtzzentrum f\"ur Schwerionenforschung GmbH, 64291 Darmstadt, Germany}
\affiliation{Max-Planck-Institut f\"ur Kernphysik, Saupfercheckweg 1, 69117 Heidelberg, Germany}

\preprint{CERN-TH-2026-275}

\begin{abstract}
Ultrarelativistic \Ox+\Ox\ and \Ne+\Ne\ collisions recently performed at the Large Hadron Collider have imaged quadrupole and octupole correlations between nucleons through the collective flow of emitted hadrons. We establish a quantitative link between these correlations and the low-energy constants (LECs) that enter chiral inter-nucleon interactions. Emulators built using the subspace-projected coupled-cluster method enable a global sensitivity analysis of the ground-state quadrupole mean-square eccentricity, revealing that quadrupole correlations respond linearly and additively to all 16 LECs at next-to-next-to-leading order in the chiral expansion. The sensitivities differ markedly in \Ox\ and \Ne, and are complementary to those of the binding energy and charge radius, the bulk observables to which nuclear interactions are typically calibrated. In particular,  \Ne\ collisions dominantly probe the spin-orbit part of two-nucleon interactions. Soft-particle correlations in light-ion collisions thus constitute a new type of data to study nuclear forces and their emergence from quantum chromodynamics.
\end{abstract}

\maketitle

\textit{Introduction.--} Measurements of soft-particle correlations in the final state of high-energy nuclear collisions \cite{Heinz:2013th,Ollitrault:2023wjk} at the Relativistic Heavy Ion Collider (RHIC) and the Large Hadron Collider (LHC) reveal imprints of the low-energy structure of the colliding ions \cite{STAR:2015mki,ALICE:2018lao,CMS:2019cyz,ATLAS:2019dct,ATLAS:2022dov,STAR:2024wgy,ALICE:2024nqd,STAR:2025elk,CMS:2025opi,STAR:2021mii,ATLAS:2025nnt,ALICE:2025luc,CMS:2025tga,ATLAS:2026zgq}, bridging TeV-scale collision dynamics with MeV-scale nuclear phenomena \cite{Jia:2022ozr}. These imprints were initially interpreted based on schematic nuclear structure models such as the classical rotor model \cite{Giacalone:2017dud,Zhang:2021kxj,Nijs:2021kvn,Zhao:2022uhl,Ryssens:2023fkv,YuanyuanWang:2024sgp,Zhao:2024feh,Giacalone:2024luz,Giacalone:2024ixe,Fortier:2024yxs,Mantysaari:2024uwn,Liu:2025zsi,Li:2025hae,Xu:2025cgx,Zhang:2025zrm,Mehrabpour:2026lhj,MenonKavumpadikkalRadhakrishnan:2026tfn,Bally:2021qys,Bally:2023dxi}. However, a paradigm shift is ongoing in which final-state observables are related to ground-state correlation functions and operators expectation values, linking collider results to nuclear structure properties emerging from a full-fledged quantum mechanical description~\cite{Giacalone:2023hwk,Blaizot:2025scr,Duguet:2025hwi,Blaizot:2025bfu,Bofos:2026huw,Bofos:2026nmg,Mehrabpour:2026yuc,Ding:2026foi,Barata:2026aak,Deng:2026wxb}. In particular the elliptic flow in ultra-central collisions reveals quadrupolar correlations in the $J^\pi=0^+$ ground state of the collided nuclei \cite{Duguet:2025hwi}. 

The recent breakthrough of light-ion collisions at LHC, with \Ox+\Ox\ and \Ne+\Ne\ collisions performed at the same nucleon-nucleon center-of-mass energy of $\sqrt{s_{\rm NN}}=5.36$ TeV, opens a particularly favorable setting to pursue this connection. The ALICE, ATLAS and CMS collaborations observed enhanced elliptic flow in central  \Ne+\Ne\ collisions relative to \Ox+\Ox~\cite{ALICE:2025luc,ATLAS:2025nnt,CMS:2025tga}, consistent with stronger quadrupole correlations in \Ne\ \cite{Giacalone:2024luz}. Both the doubly magic \Ox\ nucleus and the doubly open-shell \Ne\ nucleus are accessible to \textit{ab initio} calculations starting from nuclear interactions derived within effective field theories (EFTs) of low-energy quantum chromodynamics (QCD). The new measurements thus raise the question of what multiparticle correlations in high-energy collisions can teach us about the strong nuclear force that binds nucleons together.

How correlations between nucleons derive from the nuclear force is a question for \textit{ab initio} theory~\cite{Hergert:2020bxy,Ekstrom:2022yea}, which solves the nuclear many-body problem for a given Hamiltonian with controlled approximations. Chiral EFT ($\chi$EFT)~\cite{Epelbaum:2008ga,Machleidt:2011zz,Hammer:2019poc} with explicit pions, whose interactions are constrained by the approximate chiral symmetry of QCD, delivers two-nucleon (2N) interactions, three-nucleon (3N) interactions, and higher-body interactions in the form of a systematic expansion in the ratio of a soft to a hard scale. The unresolved short-distance physics is encoded into low-energy constants (LECs), a vector $\{\vec{\alpha}\}$ of coupling constants parameterizing the contact interactions and pion-nucleon vertices to be calibrated to data~\cite{Epelbaum:2008ga,Machleidt:2011zz}. In actual calculations, the expansion is truncated at some chosen order characterized by an expected accuracy. The next-to-next-to-leading order Hamiltonian NNLO$_{\rm sat}$~\cite{Ekstrom:2015rta}, calibrated on binding energies and charge radii of $^{14}$C and $^{16}$O, as well as on $^{22,24,25}$O binding energies, is the starting point of this work.

While currently available $\chi$EFT interactions differ in the employed chiral order, regularization scheme and optimization strategy~\cite{Epelbaum:2014sza,Reinert:2017usi,Entem:2017gor,LENPIC:2018ewt,Maris:2020qne,LENPIC:2022cyu,Huther:2019ont,Piarulli:2017dwd,Epelbaum:2019kcf,Hammer:2019poc,Hebeler2021PR_3NF,Hergert:2020bxy,Arthuis:2024mnl}, they typically describe well bulk observables of stable nuclei on which they are calibrated~\cite{Jiang:2020the,Elhatisari:2022zrb,Hu:2021trw,Scalesi:2026lja}. Nuclear interactions remain, however, uncertain, their calibration being challenging and prone to fine-tuning, as shown by nuclear saturation~\cite{Hebeler:2010xb,Simonis:2017dny,Ekstrom:2017koy,Drischler:2017wtt,Hebeler:2015hla}, properties of tin isotopes~\cite{Morris2018PRL_Sn100,Demol:2026glc,Vernik:2026tvi}, and predictions far away from nuclear stability~\cite{Stroberg2021,Hu:2026texas,Erler:2012zze,Neufcourt:2019qvd,Tarasov:2018zsp,Kondo:2023lty,Hu:2024jih,Taniuchi:2019,GarciaRuiz:2016ohj,Wienholtz:2013nya,Koszorus:2020mgn,Vernik:2026tvi}. Which data should inform the LECs, within theory errors that remain challenging to quantify~\cite{Furnstahl:2015rha,Carlsson:2015vda,Epelbaum:2014efa,Melendez:2017phj,Wesolowski:2021cni,Svensson:2021lzs,Svensson:2022kkj,Svensson:2023twt}, is an open question. Light-ion collisions offer an interesting alternative to traditional low-energy data given that they probe ground-state correlations beyond those reflected in energies and radii. But whether and for which LECs this can inform nuclear interactions has not been quantified yet.

\begin{figure}[!t]
\includegraphics[width=0.9\columnwidth]{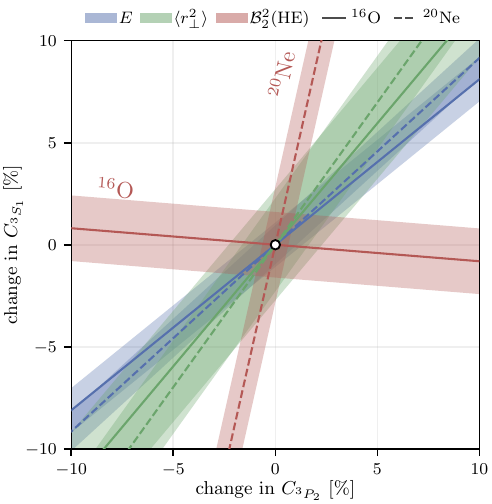}
\caption{The axes show changes in the contact LECs $C_{^3P_2}$ and $C_{^3S_1}$ relative to NNLO$_{\rm sat}$ (open circle), with the other fourteen LECs fixed. Each line marks the changes that leave one observable at its NNLO$_{\rm sat}$ value, and its band those within $\pm1\%$ (solid: \Ox, dashed: \Ne). Parallel bands mean the observables respond to the same combination of the two LECs and cannot tell them apart.}
\label{fig:strips}
\end{figure}

This work takes the necessary first step to address this question by performing a global sensitivity analysis~\cite{Sobol2001,Saltelli2010,Ekstrom:2019lss}.  Are the ground-state correlations probed by light-ion collisions sensitive to LECs other than those determining the ground-state energy and radius?  To answer this, we determine how much of the variation of the quantities probed in light-ion collisions can be explained by each of the $16$ LECs entering the NNLO$_{\rm sat}$ Hamiltonian. Figure~\ref{fig:strips} previews the result for the two LECs that matter most for quadrupole correlations: the subleading contacts in the $^{2S+1}L_J$=$^3S_1$ 2N partial wave\footnote{In the left-hand side of the equation $S$ represents the two-body intrinsic spin such that the number $2S+1$ equals 1 (spin singlet) or  3 (spin triplet). In the right-hand side S denotes the $L=0$ two-body orbital angular momentum.} for \Ox\ and in the $^3P_2$ channel for \Ne. As discussed below, we see that these LECs do affect the quadrupole correlations differently from the energy and radius, while $^{16}$O and $^{20}$Ne probe complementary combinations of LECs.


\begin{figure*}[!t]
\includegraphics[width=0.9\textwidth]{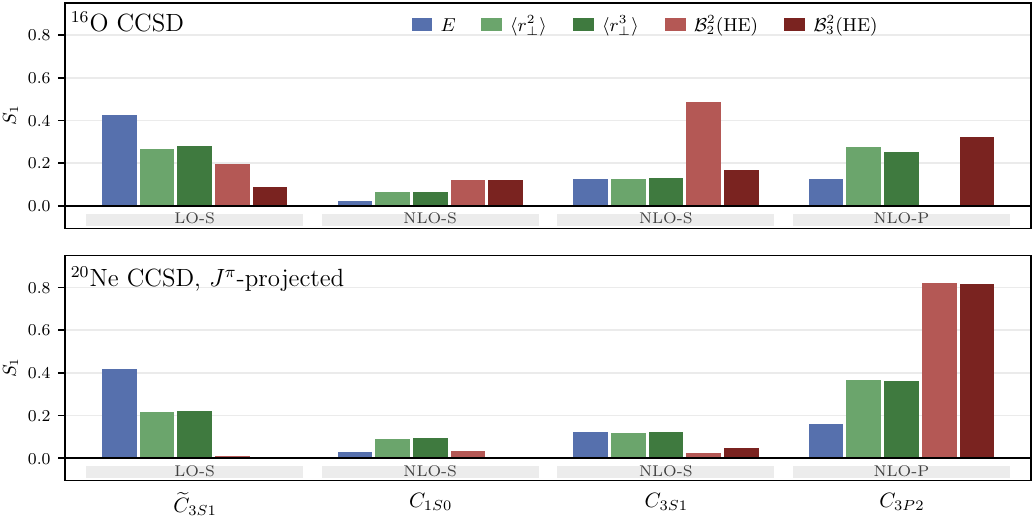}
\caption{The dominant first-order Sobol indices $S_1$ for the ground-state energy, the
transverse radii and $\Bl_\ell^2(\mathrm{HE})$ in \Ox\ (top) and \Ne\ (bottom),
against the four LECs that carry most of the variance of $\Bl_\ell^2(\mathrm{HE})$. The remaining twelve LECs carry 10--30\% of the variance together.}
\label{fig:gsa}
\end{figure*}

\textit{From detected hadrons back to nuclear forces.--}  In this work, we connect the detected hadrons to  $\chi$EFT interactions, and quantify all sensitivities. The azimuthal distribution of emitted hadrons in one collision event is expanded in Fourier harmonics as $dN/d\varphi \propto 1+2\sum_\ell v_\ell\cos[\ell(\varphi-\Psi_\ell)]$, where $\varphi$ denotes the azimuthal angle in the plane transverse to the collision axis. Since the event plane $\Psi_\ell$ cannot be located in a single event, information on the Fourier component $v_\ell$ is extracted from a two-particle correlation analysis, $v_\ell\{2\}\equiv\sqrt{\langle\!\langle\cos \ell(\varphi_1-\varphi_2)\rangle\!\rangle}=\sqrt{\langle v_\ell^2\rangle}$, involving an average over a large number of events~\cite{Borghini:2001vi,Ollitrault:2023wjk}. Through the hydrodynamic description of the quark-gluon plasma~\cite{Ollitrault:2007du,Teaney:2009qa,Heinz:2024jwu} formed shortly after the collision, the final momentum anisotropy is shown, to good approximation, to be linearly proportional to the initial spatial anisotropy~\cite{Teaney:2010vd,Gardim:2011xv,Sousa:2024msh}, i.e., $v_\ell=\kappa_\ell\epsilon_\ell$, so that $v_\ell\{2\}=\kappa_\ell\sqrt{\langle\epsilon_\ell^2\rangle}$. Here, $\epsilon_\ell$ is the spatial eccentricity of multipolarity $\ell$ of the energy density distribution deposited in the transverse plane of the reaction, which depends strongly on nuclear structure properties of the incoming nuclei, while $\kappa_\ell$ encapsulates the medium hydrodynamic response.

The eccentricities essentially reflect the distribution of nucleonic positions at the time of the collision. Consequently, their variance $\langle\epsilon_\ell^2\rangle$ over many events reveals specific, i.e., angular, two-body correlations in the ground state of the collided nuclei~\cite{Giacalone:2023hwk}. This, in turn, allows one to relate the variance over ultra-central events, i.e., events where all nucleons participate in the collision, to the quantum ground-state expectation value of a specific operator~\cite{Duguet:2025hwi}. Respectively introducing the one-body eccentricity and transverse radius operators as
\begin{equation}
  \Eop_\ell^{(1)}\equiv \sum_i(x_i+iy_i)^\ell \, ,
  \,\,
  ~~R_\ell^{(1)}\equiv\sum_{i=1}^{A}\big(x_i^2+y_i^2\big)^{\ell/2} \, ,
\end{equation}
so that $\Eop_{-\ell}^{(1)}=(\Eop_\ell^{(1)})^*$, the squared eccentricity operator of present interest is subsequently defined as $\Eop_\ell^{(2)}\equiv\Eop_\ell^{(1)}\Eop_{-\ell}^{(1)}=|\Eop_\ell^{(1)}|^2$. Eventually, the normalized mean-square eccentricity of multipolarity $\ell$ reads as
\begin{equation}
  \langle\epsilon_\ell^2\rangle
  \equiv \frac{\langle\Eop_\ell^{(2)}\rangle}{2\,\langle R_\ell^{(1)}\rangle^2}
  =\frac{\langle\Eop_\ell^{(2)}\rangle}{2A^2\langle r_\perp^\ell\rangle^2} \, ,
  \label{eq:eps}
\end{equation}
where expectation values are taken with respect to the $J=0$ nuclear ground state of the (identical) collided nuclei and where $\langle r_\perp^\ell\rangle \equiv \langle R_\ell^{(1)} \rangle /A$ denotes the standard mean nuclear transverse radius. As the operator $\Eop_\ell^{(2)}$ is a product of two one-body operators, it decomposes into the sum of a one-body operator and a two-body operator. The one-body contribution, $\langle\Eop_\ell^{(2)}\rangle^{\rm 1b}=A\langle r_\perp^{2\ell}\rangle$, gives $\langle\epsilon_\ell^2\rangle^{\rm 1b}=\langle r_\perp^{2\ell}\rangle/(2A\langle r_\perp^\ell\rangle^2)$, which scales as $1/A$. The two-body contribution delivering $\langle\epsilon_\ell^2\rangle^{\rm 2b}$ encodes two-body angular correlations of interest and is recast into a squared {\it effective} high-energy (HE) deformation parameter~\cite{Blaizot:2025bfu,Bofos:2026huw}
\begin{equation}
  \Bl_\ell^2(\mathrm{HE}) \equiv\frac{4\pi}{3}\,\langle\epsilon_\ell^2\rangle^{\rm 2b}
  =\frac{2\pi}{3}\,\frac{\langle\Eop_\ell^{(2)}\rangle^{\rm 2b}}{A^2\langle r_\perp^\ell\rangle^2} \, .
  \label{eq:B}
\end{equation}
For a classical rigid rotor (CRR) model based on a Gaussian intrinsic one-body density characterized by the axial quadrupole intrinsic deformation parameter $\beta_{20}$, one finds~\cite{Bofos:2026huw} $\langle\delta\epsilon_2^2\rangle_{\text{CRR}}=1/A+3\beta_{20}^2/(4\pi)$ to leading order in $\beta_{20}$, i.e., the one-body contribution is trivial and equal to $1/A$ whereas $\Bl_2^2(\mathrm{HE})_{\text{CRR}}=\beta_{20}^2$ (for $\ell=3$, $\Bl_3^2(\mathrm{HE})_{\text{CRR}}=1.13\,\beta_{30}^2$); i.e., the squared effective deformation reduces to the squared intrinsic deformation. Quantum corrections to the classical rotor picture have been quantified in Ref.~\cite{Bofos:2026nmg}. Many-body correlations, necessary for an accurate solution of A-body Schr\"odinger's equation (see below), further impact the value of $\Bl_2^2(\mathrm{HE})$~\cite{Ding:2026foi}.

We note that, whenever using a $J=0$ state, $\langle \Eop_\ell^{(2)}\rangle$ delivers nothing but the expectation value of the second-order Kumar-Cline invariant $\langle Q_\ell \cdot Q_\ell \rangle$~\cite{Kumar:1972zza,Cline:1986ik} built from the {\it mass} multipole operator $Q_{\ell\mu}\equiv\sum_i r_i^\ell Y_{\ell\mu}(\Omega_i)$~\cite{Bofos:2026huw}. Thus, flow observables in ultrarelativistic collisions offer a closely related and complementary perspective to low-energy Coulomb excitation experiments aiming at reconstructing the {\it charge} Kumar invariants~\cite{Bree:2014mxa,Ayangeakaa:2019psv,Henderson2019a,Rocchini2021a,Garrett2024a}. While the former reaction process offers the advantage of directly accessing the exact, i.e., complete, Kumar mass invariants, the latter is not limited to near-stable nuclei.

Collecting all the pieces, the goal is to quantify the way the measured azimuthal flow depends on the LECs $\{\vecalpha\}$ through
\begin{equation}
  v_\ell\{2\}^2 \approx \kappa_\ell^2\Big[\langle\epsilon_\ell^2(\vecalpha)\rangle^{\rm 1b}+\frac{3}{4\pi}\,\Bl_\ell^2(\mathrm{HE};\vecalpha)\Big].
  \label{eq:v}
\end{equation}
Because the one-body contribution is trivial, i.e., it remains equal to $1/A$ to a very good approximation, the present analysis focuses on $\Bl_\ell^2(\mathrm{HE};\vecalpha)$ encoding non-trivial two-body correlations. 

\textit{Global sensitivity analysis.--} The ground state $|\Psi(\vecalpha)\rangle$ depends on the LECs through the A-body non-relativistic stationary Schr\"odinger equation $H(\vecalpha)|\Psi(\vecalpha)\rangle=E(\vecalpha)|\Psi(\vecalpha)\rangle$ and, with it, every expectation value introduced above. We solve for the nuclear ground states of \Ox\ and \Ne\ using coupled-cluster theory~\cite{kuemmel1978,bishop1991,bartlett2007,Hagen14RPP} based on the singles and doubles (CCSD) approximation, which accurately describes ground-state properties with a few percent error. A given calculation yields the binding energy $E$, the transverse radii $\langle r_\perp^2\rangle$ and $\langle r_\perp^3\rangle$, as well as the effective squared deformation parameter $\Bl_\ell^2$ for $\ell=2,3$.  Since an accurate description of multipolar correlations in doubly open-shell nuclei builds on a reference mean-field state breaking rotational symmetry~\cite{Novario2020a,Frosini:2024ajq,Scalesi:2024nao,Scalesi:2026zrw,Scalesi:2026lja}, the \Ne\, CCSD ground state is further projected onto good angular momentum $J=0$ following the disentangled bi-variational coupled-cluster approach~\cite{qiu2017,Sun:2024iht}. Because parity breaking is not explored in the present work, the quantitative description of octupole correlations, eventually encoded into $\Bl_3^2$, is rather crude but can be improved in the future by relaxing such a constraint~\cite{Frosini2021mrII,Hagen:2026dqg}. 

A global sensitivity analysis with respect to $16$ LECs requires of order $10^6$ model evaluations to converge all variance integrals, which is only possible by efficiently and accurately emulating the full-space coupled-cluster calculations. This is done at a rate of $10^3$--$10^4$ evaluations per second with percent-level accuracy, via the subspace-projected coupled-cluster (SPCC) method~\cite{Ekstrom:2019lss} based on the eigenvector continuation technique~\cite{Frame:2017fah,Konig:2019adq,Duguet:2023wuh,Franzke:2025pvo,Belley:2026prl}. Parameters of the CCSD calculations, the training data for the SPCC emulators and their validation are documented in the End Matter.

Global sensitivity is quantified by the first-order Sobol index $S_i=\mathrm{Var}\,\mathbb{E}[Y|\alpha_i]/\mathrm{Var}(Y)$, the fraction of the variance of an observable $Y$ that would be removed by learning $\alpha_i$ exactly~\cite{Sobol2001,Saltelli2010}.  If $\sum_i S_i=1$, the response is additive and the first-order indices provide a complete picture.  We use a Saltelli design of $N(2k+2)=557\,056$ evaluations with $k=16$.  The LECs are varied independently in a domain centered on the NNLO$_{\rm sat}$ parameterization, with half-widths $0.005\times 10^4$ GeV$^{-2}$ for the leading order (LO) contacts $\widetilde{C}$, $0.05\times 10^4$ GeV$^{-4}$ for the next-to-leading-order (NLO) contacts $C$, $0.05$ GeV$^{-1}$ for the pion-nucleon LECs $c_i$, and $0.05$ for the 3N interaction LECs $c_D,c_E$. This domain is identical to the one used in Ref.~\cite{Sun:2024iht} and is motivated both from recent Bayesian analyses of chiral 2N interactions~\cite{Svensson:2022kkj} and from naturalness arguments. The hierarchy of the 2N posterior widths of Ref.~\cite{Svensson:2022kkj} is maintained, the width of the leading-order contacts being an order of magnitude narrower than for the higher-order ones. The resulting spread of each observable is shown in Fig.~\ref{fig:gsa_histograms} in the End Matter.

\textit{Complementary sensitivities across energy scales.--} Figure~\ref{fig:gsa} contains the main result of this work. Both in \Ox\ and \Ne, the first-order indices sum to $0.99$--$1.00$ for all five observables of interest, i.e., the response is systematically additive.  The sensitivities of energies and radii look alike in both nuclei: the leading-order $^3S_1$ contact sets the binding scale ($S_1=0.43$ and $0.42$) whereas radii mix $S$- and $P$-wave sensitivities similarly. Two-body quadrupole correlations encapsulated into $\Bl_2^2$ break this pattern in ways that differ in \Ox\ and \Ne.  In the latter case, $\Bl_2^2$ is effectively a one-parameter observable, with the $^3P_2$ contact carrying $S_1=0.82$ and no other LEC exceeding $0.04$.  In \Ox, $\Bl_2^2$ is led by the NLO $^3S_1$ contact with $S_1=0.49$, followed by the LO $^3S_1$ contact ($0.20$) and the $^1S_0$ contact ($0.12$). The $^3P_2$ main effect is very small, $S_1=0.005$. Summed by orbital angular momentum, the $\Bl_2^2$ sensitivity is $85\%$ $P$-wave and $11\%$ $S$-wave in \Ne, whereas it is $90\%$ $S$-wave and $1\%$ $P$-wave in \Ox. In \Ox, the sensitivity to $P$-wave contacts drop out of $\Bl_2^2$ because they change the two-body eccentricity and the squared radius in the same proportion. About 70\% of the $P$-wave sensitivity of $\Bl_2^2$ in \Ne\ originates from the 2N spin-orbit contact interaction, whereas the energy and radius respond to the central part, see End Matter for details.

Interestingly, quadrupole correlations in the two nuclei probe nearly independent combinations of LECs. The angle between both $\Bl_2^2$ gradients in the $16$-dimensional space, weighted by the domain half-widths, is $81^\circ$, against pairwise angles of $4^\circ$--$31^\circ$ among the four energy and radius gradients. Relative to the plane associated with $E$ and $\langle r_\perp^2\rangle$ in the same nucleus, $\Bl_2^2$ lies $38^\circ$ outside of it in \Ne\ and $54^\circ$ in \Ox. Figure~\ref{fig:strips} shows the same geometry projected onto the two LECs that matter the most: the energy and radius bands from both nuclei form one bundle, whereas the two $\Bl_2^2$ bands cross it and each other. The quoted angles depend on the chosen parameter domain and we find that $81^\circ$ reduces to $66^\circ$ when using equal relative widths. However, which gradient components vanish, and their ordering, do not depend on the domain choice. 

Repeating the \Ox\ analysis with the more sophisticated CCSDT-3 many-body truncation changes the first-order index of any observable by less than $0.07$ and leaves the leading LEC of $\Bl_2^2$ unchanged at $0.49$. We conclude that the outcome of the sensitivity analysis remains unchanged when advancing the many-body solution to even higher accuracy.

Finally, the parameter $\Bl_3^2$ shown in Fig.~\ref{fig:gsa} is dominantly sensitive to the $^3P_2$ contact in both nuclei ($S_1=0.32$ in \Ox\ and $0.82$ in \Ne). This pattern will have to be confirmed in future calculations based on the breaking and restoration of parity symmetry~\cite{Frosini2021mrII,Hagen:2026dqg}.

\textit{Conclusions and outlook.--} The present study is the first to establish how two-body quadrupolar and octupolar correlations imaged in ultrarelativistic light-ion collisions depend on the LECs of underlying $\chi$EFT inter-nucleon interactions. 

The quadrupole correlations parameter $\Bl_2^2$ responds linearly and additively to all $16$ LECs at NNLO in the $\chi$EFT expansion. In \Ox\ and \Ne, $\Bl_2^2$ probes nearly orthogonal directions, and in each nucleus its LEC-gradient lies outside the plane spanned by those of the energy and the radius. The collisions predominantly respond to the $^3S_1$ and $^3P_2$ partial waves of 2N contact interactions, and in \Ne\ specifically to the 2N spin-orbit contact interaction, which is at variance with the sensitivities displayed by energies and radii. It is instructive to further compare these results with those obtained for low-energy observables that are also sensitive to quadrupole collectivity in nuclei. The $E(4^+_1)/E(2^+_1)$ ratio in neon and magnesium isotopes, for instance, is mainly driven by the $S$-wave contacts, mostly the $^1S_0$ wave~\cite{Sun:2024iht}, which is interpreted as an effect of nuclear pairing~\cite{Scalesi:2026cpe}. The quadrupole moments of $^6$Li and $^{12}$C are also sensitive to $S$ waves~\cite{Becker:2024qdz}. This indicates that low- and high-energy probes are complementary and that their combination can improve our knowledge of inter-nucleon interactions rooted in QCD.

Regarding low-energy theory, the next steps are to quantify the impact of the many-body uncertainty and the EFT truncation error on $\Bl_\ell^2$, perform a Bayesian analysis of the LECs to quantify how data from light-ion collisions impact their uncertainties, for example via history matching~\cite{Vernon:2010,Hu:2021trw,Kondo:2023lty,Jiang:2022tzf}, and to use Bayesian experimental design to prioritize the ion collisions that are most informative regarding various nuclear structure properties of interest. For octupole correlations $\Bl_3^2$, which govern the triangular flow $v_3$, the description of \Ne\ must be extended to parity-breaking reference states. 

On the high-energy side, the challenge is to measure and construct observables that enable one to isolate the impact of two-body correlations, and thus the value of $\Bl_2^2$ for $^{20}$Ne, on top of the dominant one-body contribution, $\langle \epsilon_2^2\rangle^{1\rm b}$. This must be done by taking into account theory uncertainties on the value of $\kappa_\ell$ and the validity of the linear response approximation. A recent global Bayesian analysis of LHC data from $^{129}$Xe collisions \cite{Giacalone:2026fat} achieved the first such extraction of $\Bl_2^2$, including systematic theoretical uncertainties, within the classical rotor model. For $^{20}$Ne, such an extraction would be more challenging due to the smaller system size, but the clear signals of strong quadrupole correlations already reported in \Ne+\Ne\ collisions are promising in this respect. Further, recent results within kinetic theory indicate that the standard viscous hydrodynamic description of light-ion collisions is appropriate to robustly quantify nuclear-geometry effects from the elliptic flow of these systems \cite{Ambrus:2026ptu}.

The present paper focuses on a single observable, the variance of the azimuthal flow accessible via two-particle correlations. Multi-particle correlation techniques in high-energy collisions are, however, highly developed and enable the measurement of a wide array of observables giving access to higher-order correlations and operators, see, e.g., \cite{Mehrabpour:2026yuc} for a first analysis of three-body correlations within a triaxial rotor model. Data from RHIC and LHC are already on tape from symmetric collisions involving $^{16}$O, $^{20}$Ne, $^{96}$Ru, $^{96}$Zr, $^{129}$Xe, $^{197}$Au, $^{208}$Pb, $^{238}$U, and the exploration of more nuclear species may be possible in the next decade of LHC experiments. We are thus at the start of a broad multi-ion-species program where flow correlations are employed as probes of nuclear correlations of various natures and, through them, the LECs of chiral interactions. Such a program can shed complementary light on the origin of the nuclear force as the low-energy residual of QCD, and improve theoretical predictions far from stability.

\begin{acknowledgments}
\textit{Acknowledgements.--} 
The authors thank B.~Bally for useful comments and support in benchmarking the implementation of the eccentricity operators. The authors thank the participants of the ExtreMe Matter Institute EMMI workshop on Light Ion Collisions 2026 for inspiring discussions. G.~H. acknowledges support from a Jubilee Professorship at Chalmers University of Technology and from the U.S. Department of Energy, Office of Science, Office of Nuclear Physics, under Award No.~DE-FG02-96ER40963, and by SciDAC-5 (NUCLEI collaboration). A.~E. was supported by the Swedish Research Council Grant No. 2024-04681. Computer time was provided by the Innovative and Novel Computational
Impact on Theory and Experiment (INCITE) program.  This research used
resources of the Oak Ridge Leadership Computing Facility at Oak Ridge
National Laboratory, supported by the Office of Science of the Department of
Energy under Contract No.~DE-AC05-00OR22725. 
 A.~T. was supported by the European Research Council (ERC) under the European Union's Horizon Europe research and innovation programme (Grant Agreement No.~101162059).
 T.~M. was supported in part by JST ERATO Grant No. JPMJER2304, Japan and by JSPS KAKENHI Grant Numbers 25K07294, 25K00995, 25K07330, and 26H01394.
\end{acknowledgments}

\bibliography{bibliography.bib}

\clearpage

\section*{End Matter}
\appendix
\section{Subspace coupled cluster method}
\label{app:cc}
To construct the emulators using the subspace coupled-cluster method, we perform a set of full-space coupled-cluster calculations for LECs selected by Latin-hypercube sampling (LHS) within \(10\%\) of the NNLO$_{\rm sat}$ values \cite{Ekstrom:2015rta}, with appropriately scaled ranges for \(c_D\) and \(c_E\) and the charge-dependent \(^{1}S_0\) contact terms constrained to vary together. We use a  harmonic-oscillator basis of $11$ major shells for \Ox\ and $9$ major shells for \Ne\ with the oscillator frequency $\hbar\omega=18$ MeV. The 3N matrix elements are further truncated via the energy cut $E_{\rm 3max}=16$ and the 3N interaction is approximated at the rank-reduced two-body level using a spherical equal-filling Hartree-Fock density~\cite{Frosini21a}. The full-space CCSD calculations of \Ox\ start from the NNLO$_{\rm sat}$ Hartree-Fock reference state. For \Ne, we use an axially symmetric, prolate harmonic-oscillator reference state obtained by filling the two valence protons and neutrons in the \(j_z=\pm 1/2\) magnetic sub-states of the \(0d_{5/2}\) orbital. We choose 64 snapshots (training points) for $^{16}$O while for \Ne, a greedy search algorithm selects a $47$-dimensional subspace from these $64$ snapshots, which improves the performance of emulator. 

Once the snapshots have been selected, we construct the subspace projected norm matrix \(N\) and the Hamiltonian matrix \(H(\boldsymbol{\alpha}_{\odot})\) for a target set of LECs \(\boldsymbol{\alpha}_{\odot}\). The coefficients of the target state, \(\boldsymbol{c}(\boldsymbol{\alpha}_{\odot})\), are obtained by solving the generalized eigenvalue problem
\begin{equation}
H(\boldsymbol{\alpha}_{\odot})\boldsymbol{c}(\boldsymbol{\alpha}_{\odot})
=
E(\boldsymbol{\alpha}_{\odot})N\boldsymbol{c}(\boldsymbol{\alpha}_{\odot}).
\end{equation}
We also construct the angular-momentum-projected norm and operator kernels in the snapshot subspace,
\begin{equation}
N^{J}_{ij}
=
\langle L_i|P^{J}|R_j\rangle,
\qquad
O^{J}_{ij}
=
\langle L_i|P^{J}O|R_j\rangle,
\end{equation}
where \(|R_i\rangle\) and \(\langle L_i|\) denote the right and left CCSD training states, respectively. Here $P^J$ is the angular momentum projection operator acting only on the deformed intrinsic $^{20}$Ne state, since $^{16}$O is spherical. Using the same coefficient vector obtained from the unprojected generalized eigenvalue problem, the projected expectation value at the target LECs is
\begin{equation}
\langle O\rangle_{J=0,\boldsymbol{\alpha}_{\odot}}
=
\frac{
\boldsymbol{c}^{\dagger}(\boldsymbol{\alpha}_{\odot})
O^{J=0}
\boldsymbol{c}(\boldsymbol{\alpha}_{\odot})
}{
\boldsymbol{c}^{\dagger}(\boldsymbol{\alpha}_{\odot})
N^{J=0}
\boldsymbol{c}(\boldsymbol{\alpha}_{\odot})
}.
\end{equation}
Thus, once the reduced kernels have been constructed, evaluating a projected observable at a new target interaction requires only the solution of a generalized eigenvalue problem and matrix-vector contractions in dimensions \(64\) and \(47\) for \(^{16}\mathrm{O}\) and \Ne, respectively.

\begin{figure}[t]
\includegraphics[width=\columnwidth]{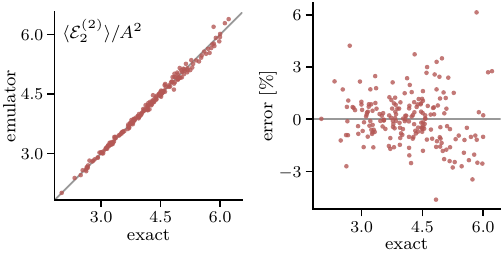}
\caption{SPCC emulator accuracy for $\langle\Eop_2^{(2)}\rangle/A^2$ in \Ne, the numerator of Eq.~(\ref{eq:eps}), computed from the $J=0$ projected CCSD solution. The one-body contribution is included.  Left: emulator against exact coupled-cluster values at $197$ cross-validation points.  Right: relative error.  The median absolute error is $0.84\%$ and the 95th percentile $2.7\%$.}
\label{fig:xval}
\end{figure}

To validate the emulators, we performed full-space coupled-cluster calculations for LEC combinations sampled by LHS from the same domain as the snapshot LECs: for \Ne, the energy, the transverse radius and, in the angular-momentum-projected state, the mean-square quadrupole eccentricity at about 200 combinations; for \Ox, the energy at 64 combinations. This domain covers the sampling ranges of the sensitivity analysis for all LECs except the leading-order contacts, for which it only covers about half the sampling ranges. Table~\ref{tab:xval} and Fig.~\ref{fig:xval} summarize the errors $|Y_{\rm SPCC}-Y_{\rm CCSD}|/|Y_{\rm CCSD}|$, where $Y$ denotes a quantity of interest. The errors on the median are below $1\%$. The error is largest for the projected $\langle\Eop_2^{(2)}\rangle/A^2$.

Finally, we note that while $\Bl_\ell^2$ values do depend on the many-body truncation, their sensitivity indices do not. For $^{16}$O, we also constructed emulators based on the more accurate CCSDT-3 approach~\cite{Noga:1987CCSDTn} which includes leading-order triples excitations.  
We found that between CCSD and CCSDT-3 the correlation of the first-order indices of the seven observables computed in \Ox\ ranges from $0.965$ to $0.998$, e.g. $0.983$ for $\Bl_2^2$, where the largest change of $0.069$ occurs for $C_{^1S_0}$.  

\begin{table}[t]
\caption{Errors of the SPCC emulators at the cross-validation points (median and 95th percentile, in per cent) and spread of the same quantity over the sampled domain (one standard deviation, in per cent of the mean).  The spread of the last row is that of the two-body part of the mean squared eccentricity.}
\label{tab:xval}
\begin{ruledtabular}
\begin{tabular}{llrrr}
 & quantity & median & 95\% & spread \\
\hline
\Ox & $E$ & $2\times10^{-6}$ & $2\times10^{-3}$ & $7.5$ \\
\Ne & $E$ & $0.10$ & $0.52$ & $8.6$ \\
\Ne & $\langle r_\perp^2\rangle$ & $0.14$ & $0.58$ & $3.6$ \\
\Ne\ proj. & $\langle\Eop_2^{(2)}\rangle/A^2$ & $0.84$ & $2.7$ & $12$ \\
\end{tabular}
\end{ruledtabular}
\end{table}

\begin{figure}[!t]
\includegraphics[width=\columnwidth]{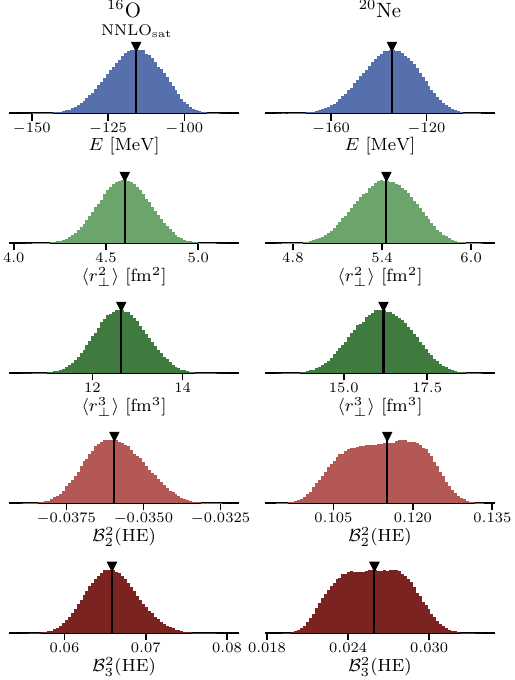}
\caption{The histograms show the spread of each observable over the sampled domain, NNLO$_{\rm sat}$ marked.}
\label{fig:gsa_histograms}
\end{figure}

\section{Origin of the quadrupole sensitivities}
\label{app:mechanism}
At fixed $A$, Eq.~(\ref{eq:B}) gives for any LEC $\alpha_i$
\begin{equation}
  \partial_i\ln\Bl_\ell^2=\partial_i\ln\langle\Eop_\ell^{(2)}\rangle^{\rm 2b}-2\,\partial_i\ln\langle r_\perp^\ell\rangle=(n_i-2)\,\partial_i\ln\langle r_\perp^\ell\rangle,
  \label{eq:slope}
\end{equation}
where $i$ labels the LEC and $n_i\equiv\partial_i\ln\langle\Eop_\ell^{(2)}\rangle^{\rm 2b}/\partial_i\ln\langle r_\perp^\ell\rangle$ is the log-log slope between the effect of $\alpha_i$ on the two-body eccentricity and on the radius.  A LEC with $n_i=2$ changes the two-body eccentricity in proportion to the radius squared and drops out of $\Bl_\ell^2$.  For $\ell=2$, every $P$-wave contact in \Ox\ has $n_i=1.5$--$2.1$ and nearly cancels, while the $S$-wave contacts that carry the sensitivity have $n_i=2.7$--$3.7$ and survive.  In \Ne\ the $^3P_2$ contact has $n_i=4.6$, far from $2$, so its effect is not cancelled by the radius.

The contact interaction at next-to-leading order~\cite{Machleidt:2011zz} sums central, spin-spin, spin-orbit and tensor-type terms.  Tensor-type terms can be decomposed into a scalar part and a traceless part given by the tensor operator
\begin{equation}
  S_{12}(\hat{\mathbf{q}})=3(\boldsymbol{\sigma}_1\!\cdot\!\hat{\mathbf{q}})(\boldsymbol{\sigma}_2\!\cdot\!\hat{\mathbf{q}})-\boldsymbol{\sigma}_1\!\cdot\!\boldsymbol{\sigma}_2,
\end{equation}
which has $\hat{\mathbf{r}}$ in place of $\hat{\mathbf{q}}$ in coordinate space. Projected onto the triplet $P$ waves $^3P_J$, which share $L=S=1$ and differ only in $J=0,1,2$, one obtains~\cite{Machleidt:2011zz}
\begin{equation}
  C_{^3P_J}=C_{\rm C}+w_{\rm LS}(J)\,C_{\rm LS}+w_{\rm T}(J)\,C_{\rm T},
  \label{eq:ls}
\end{equation}
where $C_{\rm C}$ collects central and spin-spin terms, as well as the scalar parts of the tensor-type terms, whereas $C_{\rm LS}$ and $C_{\rm T}$ denote spin-orbit and tensor components. 
\begin{table}[h]
\caption{Share of the variance carried by the central, spin-orbit and tensor operators, for independent operator coefficients that vary over domains of equal width.}
\label{tab:ls}
\begin{ruledtabular}
\begin{tabular}{llccc}
 & $Y$ & central & spin-orbit & tensor \\
\hline
\Ne\ & $\Bl_2^2$ & $0.26$ & $0.71$ & $0.03$ \\
\Ox\ & $\Bl_2^2$ & $0.77$ & $0.17$ & $0.06$ \\
\Ne\ & $E$ & $0.94$ & $0.05$ & $0.01$ \\
\Ne\ & $\langle r_\perp^2\rangle$ & $0.86$ & $0.14$ & $0.00$ \\
\end{tabular}
\end{ruledtabular}
\end{table}
The decomposition to the triplet $P$ waves: for $^1P_1$ the spin is $S=0$ such that $\mathbf{L}\!\cdot\!\mathbf{S}$ and $S_{12}$ vanish, whereas for $^3S_1$ the orbital angular momentum is $L=0$ such that $\mathbf{L}\!\cdot\!\mathbf{S}$ vanishes and $S_{12}$ only connects the state to the $^3D_1$ wave.

Since the response is linear over the sampled domain, the sensitivity of an observable $Y$ to each operator coefficient is constant and follows from the chain rule,
\begin{equation}
  \frac{\partial Y}{\partial C_k}=\sum_{J=0}^{2}w_k(J)\,\frac{\partial Y}{\partial C_{^3P_J}},\qquad k={\rm C,LS,T},
\end{equation}
with the weights $w_{\rm C}=1$, $w_{\rm LS}(J)=\langle\mathbf{L}\!\cdot\!\mathbf{S}\rangle_J=(-2,-1,+1)$ and
$$  w_{\rm T}(J)=\langle S_{12}\rangle_J=-12\,(-1)^J\left\{\begin{smallmatrix}1&1&J\\1&1&2\end{smallmatrix}\right\}=(-4,+2,-\tfrac25),
$$ where the $J$ dependence of the tensor weight is that of a $6j$ symbol~\cite{Varshalovich:1988ifq}.

For $\Bl_2^2$ in \Ne\ the gradient with respect to $(C_{^3P_0},C_{^3P_1},C_{^3P_2})$ is $(-0.095,\,-0.117,\,1)$ in units of its $^3P_2$ component, and the chain rule gives $\partial Y/\partial C_k=0.79$, $1.31$ and $-0.25$ for $k={\rm C,LS,T}$ in the same units.  For operator coefficients that vary independently over domains of equal width $w$, the linear response implies that coefficient $k$ contributes $(\partial Y/\partial C_k)^2w^2/12$ to the variance of $Y$, so that the share of operator $k$ is $(\partial Y/\partial C_k)^2/\sum_{k'}(\partial Y/\partial C_{k'})^2$.  For $\Bl_2^2$ in \Ne\ the shares are $0.71$ for the spin-orbit, $0.26$ for the central and $0.03$ for the tensor operator.  Table~\ref{tab:ls} lists the shares for all cases.

Under these assumptions, the $P$-wave sensitivity of $\Bl_2^2$ in \Ne\ is a spin-orbit effect, with $\approx70\%$ of its variance in the $P$-wave sector carried by the spin-orbit operator, while the energy and the radius respond to the central operator. The $P$-wave part of $\Bl_2^2$ in \Ox\ is small.

\end{document}